# Interwire and Intrawire Magnetostatic Interactions in Fe-Au Barcode Nanowires with Alternating Ferromagnetically Strong and Weak Segments

*Aleksei Yu. Samardak, Yoo Sang Jeon, Vadim Yu. Samardak, Alexey G. Kozlov, Kirill A. Rogachev, Alexey V. Ognev, Eunjin Jeong, Gyu Won Kim, Min Jun Ko, Alexander S. Samardak,\* and Young Keun Kim\**

A. Y. Samardak, V. Y. Samardak, Dr. A. G. Kozlov, K. A. Rogachev, Dr. A. V. Ognev, Dr. A. S. Samardak
Department of General and Experimental Physics,
Far Eastern Federal University (FEFU), Vladivostok 690922, Russia
E-mail: samardak.as@dvfu.ru

Dr. Y. S. Jeon
Center for Hydrogen·Fuel Cell Research
Korea Institute of Science and Technology, Seoul 02792, Republic of Korea

E. Jeong, G. W. Kim, Dr. M. J. Ko, Prof. Y. K. Kim
Department of Materials Science and Engineering
Korea University, Seoul 02841, Republic of Korea
E-mail: ykim97@korea.ac.kr

Metallic barcode nanowires (BNWs) composed of repeating heterogeneous segments fabricated by template-assisted electrodeposition can offer extended functionality in magnetic, electrical, mechanical, and biomedical applications. We can consider such nanostructures as a three-dimensional system of magnetically interacting elements with magnetic behavior strongly affected by complex magnetostatic interactions. This study discusses the influence of geometrical parameters of segments on the character of their interactions and the overall magnetic behavior of the array of BNWs having alternating magnetization. By controlling the applied current densities and the elapsed time in the electrodeposition, we regulate the dimension of the Fe-Au BNWs. We show that the Fe and Au segments are made of Fe-Au alloys with high and low magnetization. We demonstrate that the influence of the length of magnetically weak Au segments on the interaction field between nanowires is different for samples with magnetically strong 100- and 200-nm-long Fe segments using the first-order reversal curve (FORC) diagram method. With the help of micromagnetic simulations, we discover and analyze the three types of magnetostatic interactions in the BNW arrays. As a result, we demonstrate that the dominating type of interaction depends on the geometric parameters of the Fe and Au segments and the interwire and intrawire distances.

## 1. Introduction

Magnetic memory devices, based on densely packed magnetic nanoelements, could be a breakthrough in computational technologies due to their undeniable advantages, such as non-volatility, economic efficiency, speed, and endurance.[1, 2] Successful creation of such devices requires the solution of several inherent problems, one of which is determining the influence of magnetostatic fields created by magnetic elements on their neighbors.[3] Since this influence in the array can extend over distances up to 100 μm,[4] it could drastically affect the magnetic properties of one memory cell and a whole device. For example, under the influence of magnetic interactions, a magnetic memory cell with desirable 0 and 1 states, separated by an energy barrier, can change switching fields, which is critical for information writing, or even launch an avalanche-like switching of magnetization erasing of the written information. To avoid such unwanted behavior or to turn it to our benefit, one should carefully analyze magnetostatic interactions in such systems. However, at this moment, the complexity of 3-dimensional nanostructures makes it impossible to use one model considering all possible cases. Therefore, each specific magnetic system requires an investigation of a magnetostatic field distribution to define patterns that will help decrease or even negate the influence of magnetic interactions on the properties of individual memory cells.

To couple a 3-dimensional nanostructure and magnetic geometry, we employed a barcode nanowire (BNW) composed of two magnetic segments with alternating compositions. In particular, ferromagnetic BNWs have been proved to have great potential for multiple tasks based on their unique magnetic properties,[5-8] such as agents for cancer treatment,[9, 10] biomedical applications,[11, 12] and high-density 3D magnetic memory devices.[13-15] Supposing the applied potential, we could construct BNWs in the specific assembly within the array architecture. We can consider such systems as magnetic elements distributed in a 3-dimensional space, where each ferromagnetic segment acts as a source of a magnetostatic field affecting neighboring segments and adjacent nanowires.

This study designed and investigated magnetic behavior and magnetostatic interactions in Fe-Au BNW arrays. These materials were uniformly electrodeposited in nanoporous anodized aluminum oxide (AAO) membranes from a single bath,[16] depending on segment lengths. For a comprehensive study, we carried out a complex investigation of the geometrical parameters of the alumina matrix and BNWs using scanning electron microscopy (SEM). The elemental composition of each segment was studied by dispersive X-ray analysis (EDX), which showed that in both Au and Fe segments, a ferromagnetic alloy was formed with a different ratio of elements depending on the deposition conditions. A first-order reversal curve (FORC) diagram analysis[17-22] showed a nontrivial distribution of interaction fields in the samples, heavily dependent on the geometrical parameters of magnetic segments. We implemented a micromagnetic model using the MuMax$^3$ software[23] to explain such behavior and determine the magnetization configuration in the Fe-Au BNWs. This model justified its results by comparing them with experimental data of magnetic force microscopy (MFM) and integral hysteresis loops. While others usually simulated BNW arrays synthesized from one bath as cylinders divided by non-magnetic spacers,[5, 8] we approached them as a set of nanowires with the alternating magnetization in ferromagnetically weak and strong Fe-Au segments. The simulation showed a vortex domain configuration in BNWs. It revealed three types of magnetostatic interaction between opposite poles of segments, each drastically dependent on the geometrical parameters of the sample. Our results demonstrated that with an increase in the length of the ferromagnetically weak segment, the interaction between the ferromagnetically strong segments in a single BNW decreased rapidly. In contrast, the switching-induced interaction between nanowires increased.

## 2. Results

### 2.1. Microstructural and Compositional Characterization

We prepared BNW arrays with different segment lengths ($L$) to examine the effect of the segment length on the magnetic properties. We controlled the magnitude of $L$ by the duration of the pulses with the required reduction of current densities: Fe segment – $L_{Fe}$ = 100 and 200 nm for 6 and 12 s, respectively; Au segment – $L_{Au}$ = 30, 50, 100, and 200 nm for 16, 27, 54, 107 s, respectively. We named the samples after the lengths of the corresponding segments, for example, Fe(100)Au(30) for $L_{Fe}$ = 100 nm and $L_{Au}$ = 30 nm. The average total length of the nanowires is 7.32 μm for Fe(100)Au(x) series samples and 10.47 μm for Fe(200)Au(x) samples. We focused on the regulation of each series with the same repetition numbers, though the total length of Fe(200)Au(x) series becomes larger than the other. The above allows us to make a direct comparison of a pair of samples, such as Fe(100)/Au(30) and Fe(200)/Au(30), because the number of cycles has been recognized as a more critical factor affecting the possible magnetostatic interactions than the total length in the barcode system.[24,25]

We monitored the composition-dependent morphological changes of the BNW arrays embedded in the nanoporous AAO template using fractures of the samples. We observed the sharp borders between the Fe and Au segments (Figures 1 (a)–(h)). The BNW diameter (D) measurement revealed approximately 204.81±14.69 nm for all samples. We determined the average distance between the pores in the array as $d_{ip}$ = 270±30.3 nm from the top view of the nanoporous template (Figure S1). Statistical analysis of the segment lengths of the prepared samples showed their high uniformity (Figure 1 (i)). Table S1 in the Supporting Information displays the detailed length scales.

The SEM EDX analysis of the BNWs confirmed the different elemental compositions of the Fe and Au segments (Figure 2). However, it showed that both segments do not consist merely of the corresponding elements, but they were made of a Fe-Au alloy with different proportions of elements. Statistical analysis of the composition of the segments, performed for the whole set of samples, showed that the Fe segment, prepared with the current density $\rho_i = 10$ mA cm$^{-2}$, consisted of 90±2 at.% of Fe and ~10±2 at.% of Au, hereafter referred to as the Fe segment. The Au segment, synthesized with $\rho_i = 0.5$ mA cm$^{-2}$, contained ~55±1 at.% of Fe and ~45±1 at.% of Au, hereafter referred to as the Au segment. We observed a similar elemental composition of the Fe and Au segments for the entire series of samples.

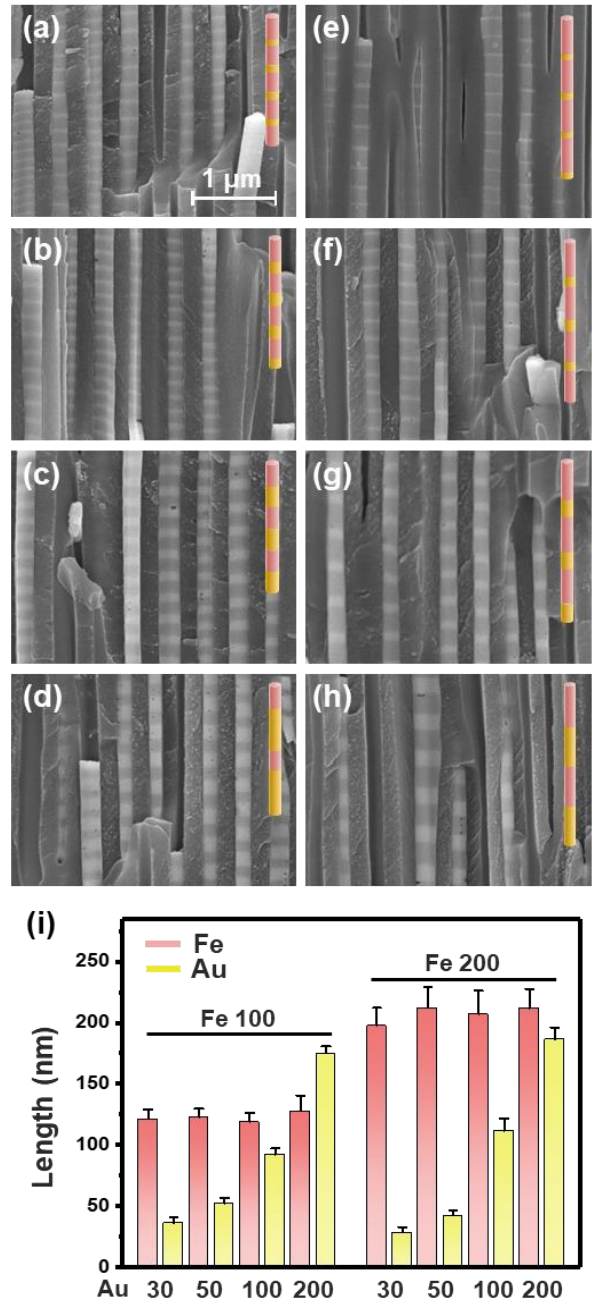

**Figure 1.** SEM images of BNWs embedded in the AAO template. They represent the following segment combinations: (a) Fe(100)Au(30), (b) Fe(100)Au(50), (c) Fe(100)Au(100), (d) Fe(100)Au(200), (e) Fe(200)Au(30), (f) Fe(200)Au(50), (g) Fe(200)Au(100), and (h) Fe(200)Au(200). (i) Length distribution of each segment for the corresponding combinations of Fe and Au segments; error bars represent the standard errors where the number of BNWs for each combination is 50 (total $n = 400$, $p = 0.05$)

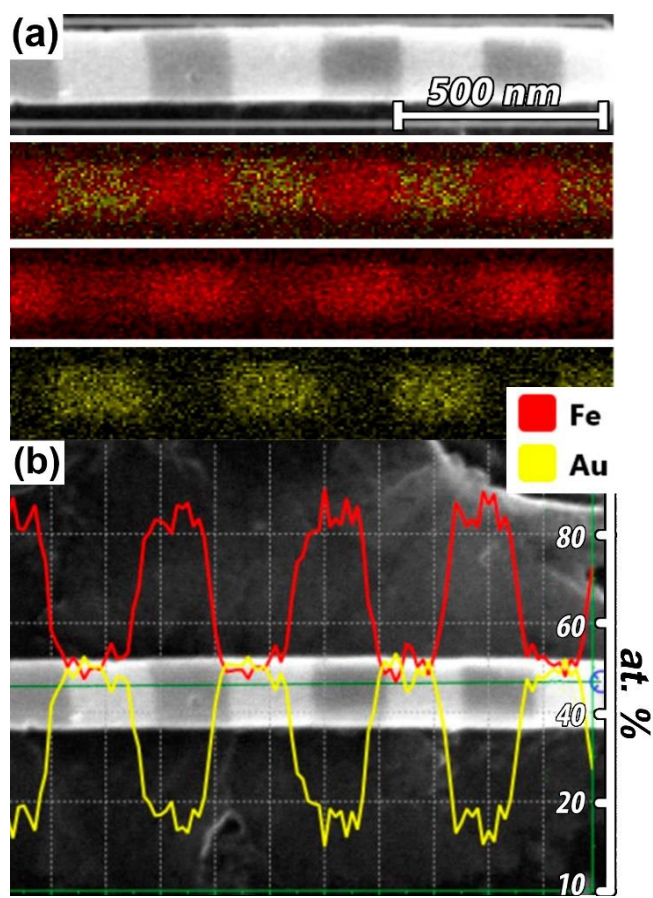

**Figure 2.** (a) SEM image and the corresponding EDX mapping, (b) line analysis of the Fe(200)Au(200) sample. The red and yellow lines represent the data characteristics for the Fe and Au atoms, respectively.

The XRD patterns acquired from all samples were indexed with PDF numbers 87–0722 and 04–0784, corresponding to the Fe and Au segments (Figure S2 (a) in the Supporting Information). We defined that the Fe and Au segments had typical body-centered cubic (bcc) and face-centered cubic (fcc) structures, respectively. The selective area electron diffraction (SAED) patterns collected from the Fe and Au segments (Figure S2 (b) in the Supplementary file) matched well with the XRD indices (Figures S2 (c), (d) in the Supporting Information). Although Au segments appeared to have a higher crystallinity, both were polycrystalline. The size of crystallites in the Au (111) phase was evaluated at approximately 12.8 nm using the Scherrer equation. Since we did not observe significant microstructural differences in all samples, we deduced that morphological and geometrical changes would be the main factors in determining the magnetic properties and micromagnetic structure of BNWs.

## 2.2. Magnetic Properties

The $Fe_{90}$-$Au_{10}$ and $Fe_{55}$–$Au_{45}$ alloys composing the Fe and Au segments are ferromagnetic.[26] According to McGuire et al.,[27] the saturation magnetization ($M_s$), depending on the weight ratio of Fe–Au, decreases linearly with an increase in the Au content. Additionally, nanowires synthesized under conditions for the Au segment showed weak ferromagnetism by vibrating sample magnetometry (VSM). Attempts to experimentally determine the saturation magnetization $M_s$ of BNWs (Figure S3 in the Supporting Information) resulted in the approximation of possible $M_s$ values in the range 800-900 emu/cm$^3$ and 200-300 emu/cm$^3$ for the Fe and Au segments, respectively. Nevertheless, this method of $M_s$ determination is rough since it depends on accurately determining the magnetic material volume in the whole sample, which is a nontrivial task for the nanowire arrays.

Figure 3 presents the hysteresis loops measured at RT for all samples. Although nanowires usually demonstrate strong uniaxial anisotropy of magnetic properties, due to the shape anisotropy of an infinitely elongated cylinder,[20] our samples in some cases showed almost isotropic magnetic behavior, with low coercive force and near-zero remnant magnetization. As one can see, Fe(100)Au(x) samples (Figure 3 (a)) are practically isotropic, with minor differences in magnetic behavior depending on the direction of an external magnetic field ($H$). The weak anisotropy of the Fe(100)Au(30) sample with the easy axis of magnetization aligned along the main axis of BNWs almost vanishes with increasing $L_{Au}$. The most pronounced anisotropy in the Fe(200)Au(30) sample is much more visible. It manifests itself in a clear difference in the hysteresis loops measured in different directions of the $H$ (Figure 3 (b)). The magnetic anisotropy decreases with increasing $L_{Au}$ in the case of the Fe(200)Au(x) series, but it is still more pronounced than in Fe(100)Au(x) samples.

The influence of two factors could explain this. The first factor is the existence of uncompensated magnetic flux on the boundaries between Fe and Au segments, resulting in dipole interactions between neighboring Fe segments. The interaction energy $U$ between two dipole moment vectors $\overrightarrow{\mu_1}$ and $\overrightarrow{\mu_2}$ with the radius-vector $\vec{r}$ between them could be written as

$$U = -\frac{\overrightarrow{\mu_1} \cdot \overrightarrow{\mu_2}}{r^3} + \frac{3(\overrightarrow{\mu_1} \cdot \vec{r})(\overrightarrow{\mu_2} \cdot \vec{r})}{r^5} \qquad (1)$$

According to Equation (1), $U$ is minimal in the case when the orientations of $\overrightarrow{\mu_1}$, $\overrightarrow{\mu_2}$ coincide and are parallel to $\vec{r}$. In that way, for intrawire interactions, the co-directional orientation of magnetic moments in adjacent Fe segments in the same BNW is energetically favorable. In the case of the interwire interactions between Fe segments in neighboring BNWs, $\overrightarrow{\mu_1}$ and $\overrightarrow{\mu_2}$ are perpendicular to $\vec{r}$ and the minimal $U$ is achieved in the opposite direction of $\overrightarrow{\mu_1}$ and $\overrightarrow{\mu_2}$. Thus, the intrawire interactions keep the co-directional magnetization in adjacent Fe segments, resulting in a more pronounced easy axis of magnetization in this direction in samples with $L_{Au}$ < 50 nm. With increasing distance between Fe segments, the dipole interactions decrease considerably. In samples with $L_{Au}$ > 50 nm, the intrawire interactions become comparable to the interwire coupling, leading to a more isotropic magnetic behavior of the BNW arrays.

The second factor is the difference in the magnetostatic energy ($E_{ms}$) of the 100- and 200-nm Fe segments due to the different geometry and demagnetizing factors ($N$). $E_{ms}$ of Au segments can be counted as negligible since it changes as the square of the $M_s$:

$$E_{ms} = \frac{1}{2}NM_S{}^2 \qquad\qquad\qquad\qquad\qquad (2)$$

As a result, it should be about 10-20 times lower than Fe segments due to the significant difference in their $M_s$. For the Fe segments, $N$ and, therefore, $E_{ms}$ will depend on the aspect ratio, $\gamma = L_{Fe}/D$. For example, for $L_{Fe} = 100$ nm ($\gamma \approx 0.4$), the demagnetizing factor along the main axis of the cylinder is $N_z \approx 0.5$ and becomes significantly lower for $L_{Fe} = 200$ nm ($\gamma \approx 0.8$) - $N_z \approx 0.36$.[28] With smaller $E_{ms}$, lesser energy of the external magnetic field ($E_{Zeeman}$) is needed to saturate the sample in the corresponding direction, making it more energetically favorable with lower saturation fields ($H_s$).

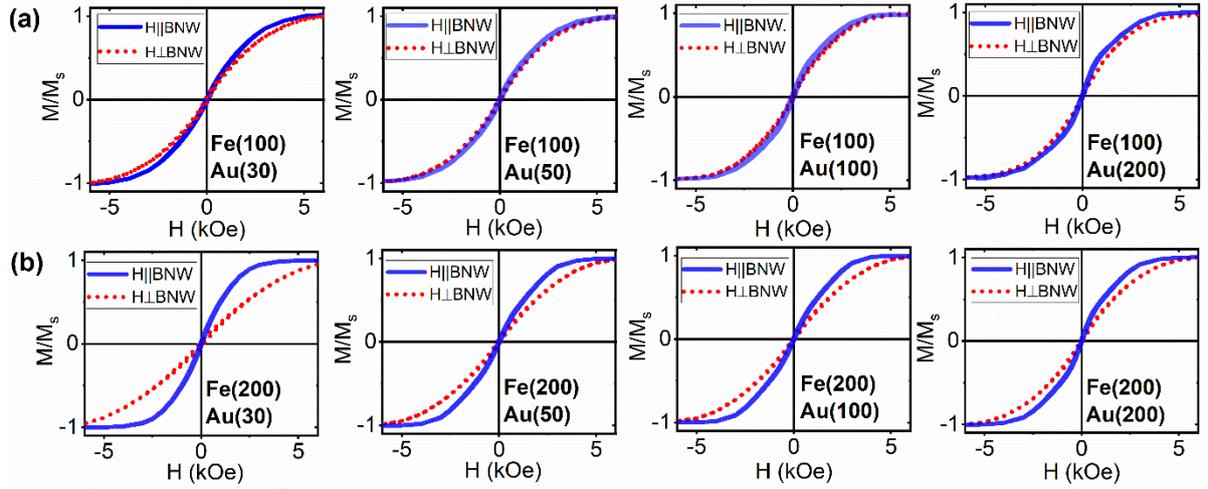

**Figure 3.** Magnetic hysteresis loops in the direction of $H$ along the main axis of BNW arrays (blue line) and perpendicular (red line) for (a) Fe(100)Au(x) and (b) Fe(200)Au(x) samples.

An inflection on the hysteresis loops measured along the main axis of BNWs in the region near 500 Oe becomes more visible with increasing $L_{Au}$, separating the regions with different slopes of the loop. The bend is most pronounced and can be visible in the hysteresis loop of the sample Fe(100)Au(200) (Figure 3 (a) and Figure S4 in the Supporting Information) since this sample has the largest aspect ratio of Fe and Au segment phases. This inflection may indicate the simultaneous magnetization process of the Fe and Au segments of two different magnetic phases. Its manifestation is associated with an increase in the amount of the Au phase in the samples.

## 2.3 FORC-diagram analysis

We carried out the FORC-diagram studies for all samples in two configurations of $H$ – along and across the main axis of BNWs. We measured in the $H$ ranging from +6000 to –6000 Oe with the 200 Oe step. In contrast to magnetic hysteresis loops, the FORC diagrams, presented in Figure 4(a–d), differ greatly depending on the direction of the applied field. The FORC diagrams plotted along the main axis of BNWs for all samples show the typical distribution of switching processes along the $H_u$ axis induced by the interaction fields (Figure 4(a, c)). In the perpendicular direction of $H$, the samples were characterized mainly by a reversible magnetization switching process, with a narrow region of irreversible switching occurring in small fields (Figure 4(b, d)).

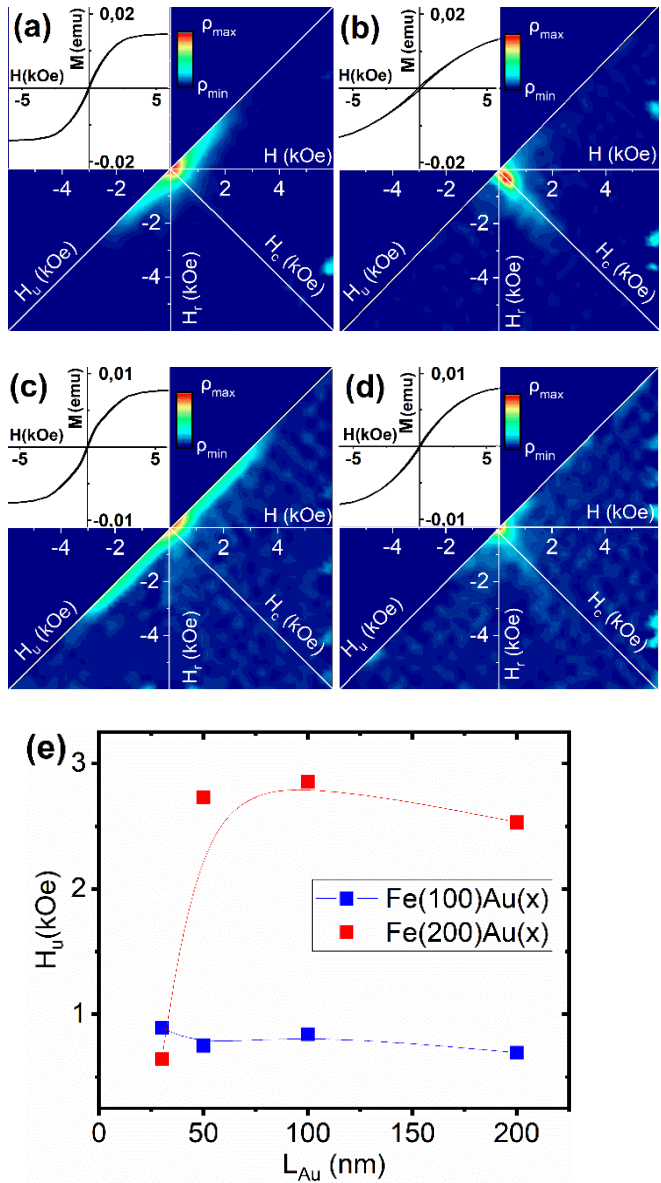

**Figure 4.** FORC diagrams for (a, b) Fe(200)Au(30) and (c, d) Fe(200)Au(200) samples measured in the $H$ along (a, c) and across (b, d) to the main axis of BNWs. The insets represent a family of experimentally measured minor curves. (e) The interaction field $H_u$ as a function of $L_{Au}$ during the interaction-induced switching for the Fe(100)Au(x) and Fe(200)Au(x) samples.

As the FORC-diagram method revealed, irreversible processes caused by magnetostatic interaction occur mostly when the $H$ is directed along the main axis of the BNWs. Otherwise, when $H$ is applied across to BNWs, a reversible magnetization switching prevails. In the field applied along BNWs, the distribution of the interaction-induced switching $\rho$ along the $H_u$ axis broadens for Fe(200)Au(x) samples in the FORC diagrams (Figure 4 (e)) with increasing $L_{Au}$. On the contrary, the $H_u$ distribution remains at the same level for Fe(100)Au(x) series. We scrutinized the micromagnetic structure of BNWs with the assistance of micromagnetic simulations to understand such an extraordinary behavior.

## 2.4 Micromagnetic structure of BNW

We chose the Fe(200)Au(200) sample as the sample with the largest segment size, making it possible to more accurately bound the structure of the BNW with the magnetostatic fields generated by it. The BNWs were etched from the AAO template and placed on a Si substrate. Using SEM, we found a single BNW with acceptable structural parameters on the substrate. Then, two orienteers were placed on the Si surface near the BNW by a focused ion beam (FIB) to find the desired nanostructure. We additionally investigated the corresponding BNW by SEM (Figure 5 (a)) for specific geometrical values such as diameter and segment lengths to make a precise simulation. Finally, we examined EDX to compare the MFM images with the segment composition (Figure 5 (b)).

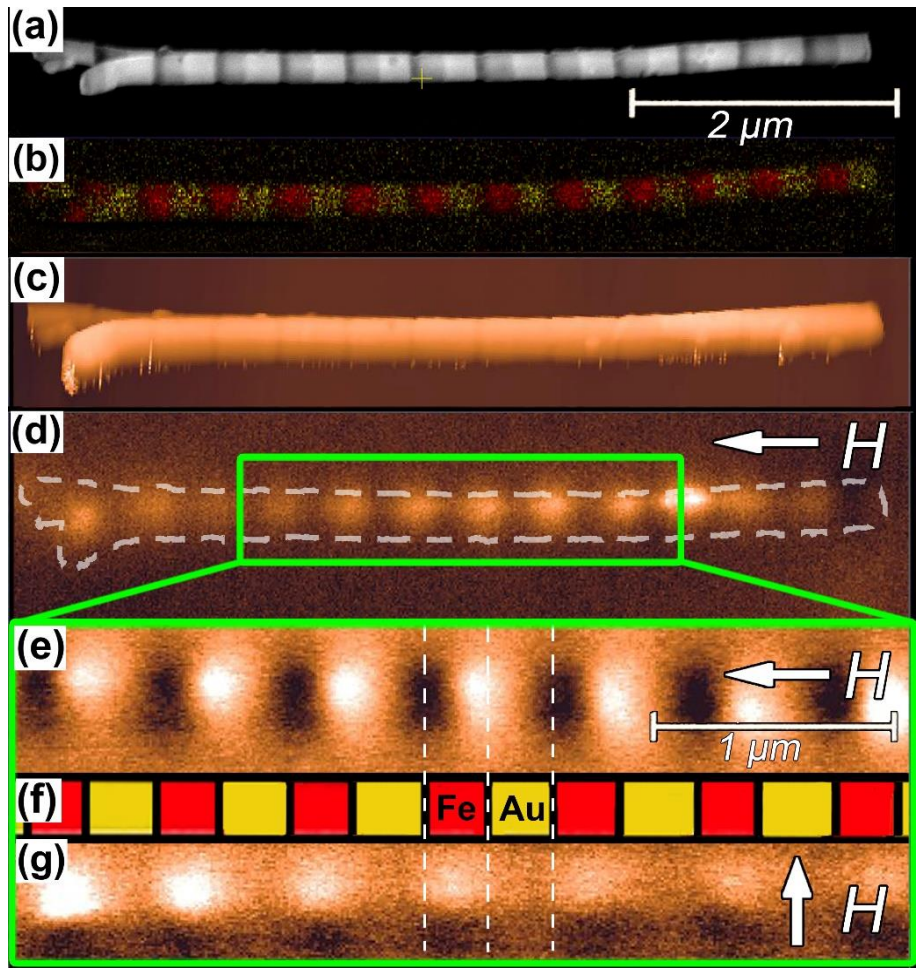

**Figure 5.** Investigation of the micromagnetic structure of the BNW: (a) SEM image; (b) elemental mapping; (c) atomic force microscopy (AFM) picture of the BNW morphology; (d) overview picture of the micromagnetic structure of the BNW in $H$=700 Oe along the main axis of the BNW, recorded in the MFM regime, where the white dashed line shows the contour of the BNW and the arrow indicates the direction of the external magnetic field; (e) detailed investigation of the part of BNW at $H = 700$ Oe along the BNW; (f) schematic image of the corresponding segments, drawn based on the EDX results, where red color exhibits the Fe segment, yellow – Au; (g) MFM contrast at $H = 700$ Oe perpendicular to the BNW.

Figure 5(c) shows an AFM image of the BNW. In the absence of $H$, we did not observe the MFM contrast for remnant states, possibly due to the vortex micromagnetic configuration in the segments, which do not create stray fields strong enough to be detected with MFM. When the longitudinal $H = 700$ Oe was applied, an alternating contrast of black and white poles became visible in the MFM picture (Figure 5(d, e)). When the distribution of the elemental composition on the MFM images (Figure 5(f)) was superimposed, it was revealed that black and white contrast is formed strictly at the boundaries between the Fe and Au segments. In the $H = 700$ Oe perpendicular to the main axis of the BNW, positions of the black and white regions rotated to 90° and were located exclusively in the Fe segments at the edges of the nanowire (Figure 5(g)), because of the weak magnetic moment of the Au segment compared to the Fe segment. In the external magnetic field ($H = 700$ Oe), because of the gradient of the magnetic moment, an uncompensated magnetic flux was formed, resulting in the formation of stray fields of different polarity on the opposite boundaries of Fe segments.

## 2.5 Micromagnetic simulations

In the simulation, the BNW array was restricted by a space of $1.5 \times 1.7 \times 2$ µm, containing 12 nanowires of hexagonal order, 2 µm long and 250 nm in diameter. Periodic boundary conditions (PBC) were implemented at the edges of the simulated area and at both ends of the BNWs to avoid boundary effects. We chose the cell size of 5 nm to be close to the length of the ferromagnetic correlation of Fe. To make the segmented structure, we divided the nanowires into regions with different values of the $M_s$ and the exchange stiffness constant ($A_{ex}$). We have tried a range of different $M_s$ values in simulations, yet only $M_s = 1000$ emu/cm$^3$ for Fe segments and $M_s = 200$ emu/cm$^3$ for Au segments showed a good overlap of simulated results with experimental data. Due to the proximity of these values to obtained results (800-900 emu/cm$^3$ for the Fe segment and 200-300 emu/cm$^3$ for the Au segment) and data from the literature for bulk FeAu alloys (1200 emu/cm$^3$ for $Fe_{90}Au_{10}$ and 600 emu/cm$^3$ for $Fe_{55}Au_{45}$ alloys), it was decided that these values are trustworthy. Therefore, for Fe segments, $A_{ex}$ was set as $15\times10^{-17}$ erg/cm; for Au segments $-A_{ex} = 5\times10^{-17}$ erg/cm. We chose the lengths of these regions close to the lengths of the Fe and Au segments, determined experimentally (Figure 1 (i)).

A comparison of MFM contrast and simulated magnetostatic fields ($B_{demag}$) is shown in Figure S5(a–e) of the Supporting Information. Experimentally measured and simulated hysteresis loops for Fe(200)Au(200) in two directions of the applied field are represented in Figure S5(f) of the Supporting Information. The simulated results show a good overlap with the experimental data. Furthermore, we achieved the same agreement between the simulated and experimental hysteresis loops for all samples.

We observed a magnetic vortex configuration in the remnant state after saturation along the main axis (Figure S5 in the Supporting Information). Although magnetization vectors in the vortex core are oriented in the direction of the previously applied field for all samples, the diameters of the cores in the Fe(200)Au(x) series are slightly larger than in Fe(100)Au(x) samples, possibly under the influence of the shape anisotropy. In the Au segments, the magnetization vectors also remain turned in the direction where the sample was saturated, twisting slightly along the edges following the vortices in the Fe segment. In the remnant state, after magnetization in the perpendicular direction of the field, the domain structure remains almost the same, with a greater twist of the magnetization in the Au segment.

The simulation showed that the switching process in the BNW occurs by the same mechanism for all samples. In the *H* applied along the main axis of BNWs, switching begins with the generation of a vortex at the boundaries between segments, in the volume of the Fe segment (see Supporting Movies 1-4 for Fe(100)Au(30), Fe(100)Au(200), Fe(200)Au(30) and Fe(200)Au(200) samples, respectively). With a subsequent decrease in the strength of the external magnetic field, the vortex spreads over the entire segment. The same process occurs in the Au segment in lesser $H$, with the beginning of vortex nucleation at the inflection point in the hysteresis loop. As mentioned above, the vortex cores remain oriented in the direction specified by the initial saturation of the nanowires and determine the small remnant magnetization of the arrays. When the critical fields are reached, the vortex cores switch following the $H$, accompanied by consistent rotation of the magnetization, until the sample is completely saturated in the direction of the external field. In Fe(x)Au(200) samples, some form of a skyrmion-like structure[29] is observed right before the vortex core switching ($H \approx \pm 1100$ Oe), with magnetization in the vortex core and at its edges oriented almost in opposite directions. Under the external magnetic field applied perpendicularly to the main axis of BNWs, magnetization reversal of Fe segments occurs through the displacement of the vortex core and its annihilation at the edge.

As the FORC-diagram analysis shown in Figure 4(e) for the Fe(100)Au(x) series, the interaction field $H_u$ stands the same regardless of the $L_{Au}$. In comparison, for the Fe(200)Au(x) samples, $H_u$ increases significantly with increasing $L_{Au}$. To explain such complex interactions in an array of BNWs, we simulated the system of two magnetized in opposite directions BNWs spaced apart by 200 nm. We simulated the samples with $L_{Fe} = 50$–400 nm and $L_{Au} = 30$–1000 nm to analyze the behavior of their magnetostatic interactions.

In our model, two nanowires were placed against each other without a shift along their length to simplify the model and facilitate calculations, although the SEM images for some samples showed the displacement of the segments on up to 7-19% of the $L$ value (Figure 1). To take into account the influence of such shift on magnetostatic fields, we simulated a system of two Fe(200)Au(200) nanowires shifted against each other in the range from 0 to 200 nm (see animation in the gif file and the derived data in Figure S7 in SI). This simulation revealed that the type I and type II interactions were slightly growing with the increase of the segment shift, while the type III interaction could drop up to 20% at the maximum shift.

As a result, we observed three possible types of interactions between the Fe segments in the BNW arrays: type (I) – intrawire interaction between the poles of adjacent Fe segments, type (II) – interaction between poles of the same Fe segment, type (III) – interwire interaction between closest poles in Fe segments of the neighboring nanowires in the array. The simulation of the magnetostatic field $B_{demag}$ and schematic representation of three types of interaction are shown in Figure 6 for each sample with the dominating type, depending on the segment geometry and interwire distance of Fe and Au.

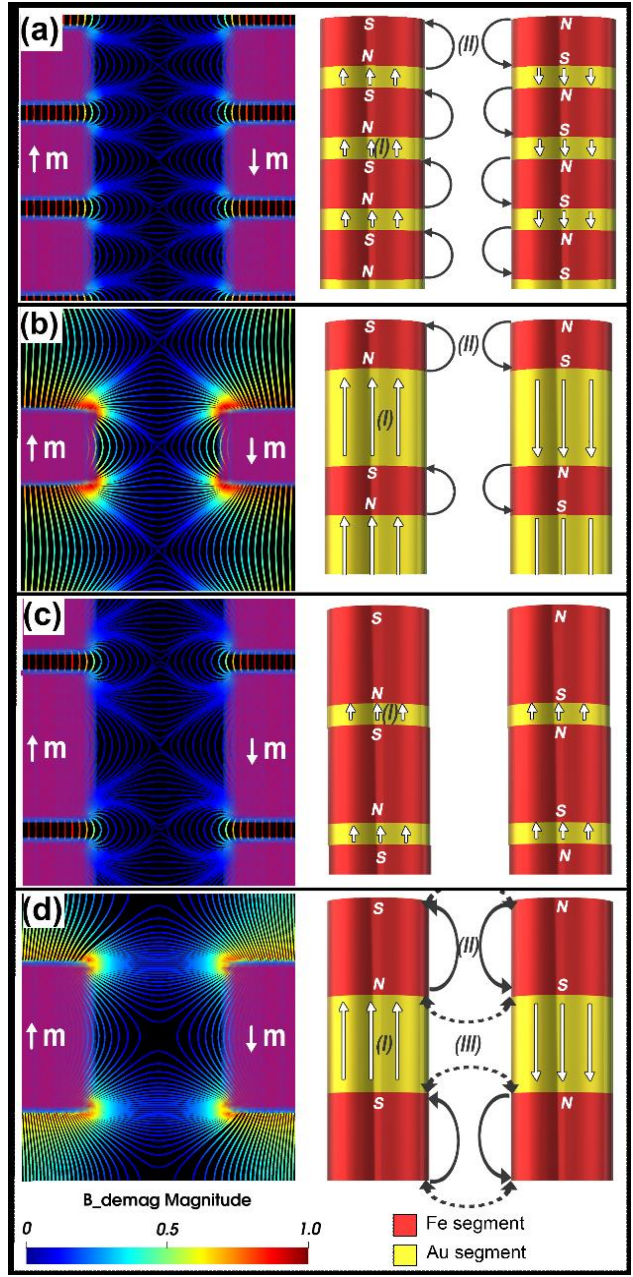

**Figure 6.** Simulation of the $B_{demag}$ distribution in two oppositely magnetized BNWs (left) and schematic representation of them (right) for (a) Fe(100)Au(30), (b) Fe(100)Au(200), (c) Fe(200)Au(30) and (d) Fe(200)Au(200) samples. On the left, the geometry of the Au segment was removed for better clarity of the image. On the right, the arrows represent the dominating type of interaction: type (I) – coupling of opposite poles in adjacent segments of the same nanowire (white arrows); type (II) - coupling of poles in the same segment (solid black arrows); type (III) - the interaction of segments in neighboring nanowires (black dashed arrows).

The results of quantitative analysis of three types of interactions in the Fe(100)Au(x) and Fe(200)Au(x) series are presented in Figures 7(a) and (b). The type (I) interaction decreases rapidly with increasing $L_{Au}$, yet it is still present in the samples with the longest Au segment (Fe(x)Au(1000)) simulated. On the contrary, the interaction of the type (II) increases with $L_{Au}$ and is much larger for samples of the Fe(100)Au(x) series because of the proximity of the poles in the Fe segments, which have the geometry of flat disks. The interaction of the type (III) grows from near-zero values for Fe(x)Au(30) samples to saturation for $L_{Au} \approx 400$ nm. Furthermore, for samples from the Fe(200)Au(x) series, the interaction grows much faster and reaches double values compared to Fe(100)Au(x). The type (I) interaction dominates in the samples with $L_{Au} < 100$ nm for Fe(100)Au(x) and $L_{Au} < 150$ nm for Fe(200)Au(x). In samples with $L_{Au}$ exceeding these values, different types of interaction dominate depending on $L_{Fe}$: the type (II) interaction in Fe(100)Au(x) and the type (III) interaction in Fe(200)Au(x) samples.

Further analysis of the distribution of demagnetizing fields in the systems of simulated BNWs with $L_{Fe} = 50–400$ nm and $L_{Au} = 30–1000$ nm resulted in a parametric diagram of the dominating type of interactions (Figure 7 (c)). As one can see, the type (I) interaction dominates mostly in the arrays with rather small Au segments ($L_{Au} < 200$ nm). In systems with $L_{Au}$ exceeding this value, the dominating type of interaction depends on $L_{Fe}$, with the type (II) dominating in samples with $L_{Fe} < 150$ nm and type (III) dominating in samples with larger Fe segments. Transition regions, where the type (III) overtakes the other two types of interactions, clearly depend on interwire distance in the array ($d_{ip} = 200$ nm) and could be easily shifted by manipulating this parameter in the alumina matrix synthesis.

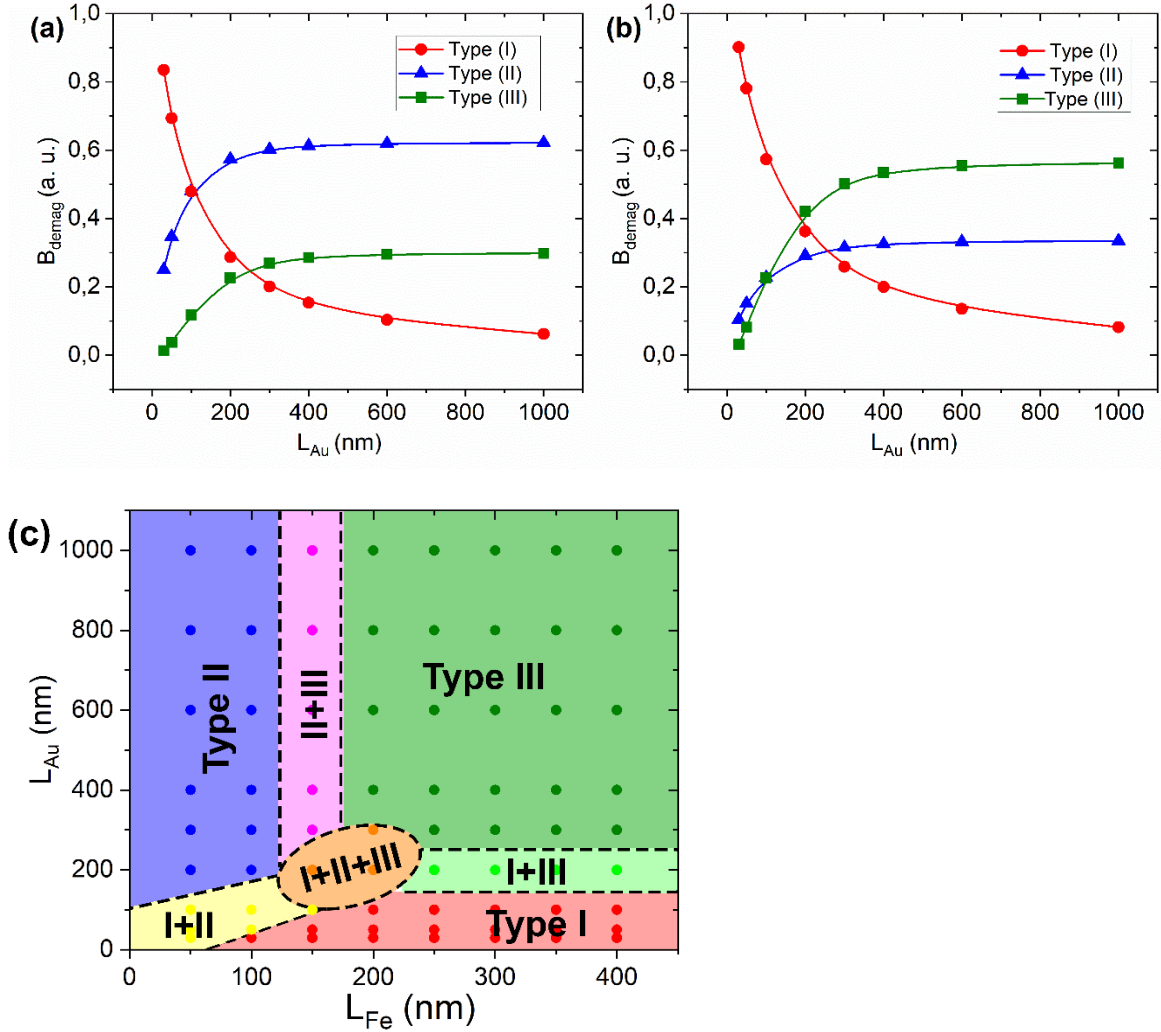

**Figure 7.** Quantitative results of the micromagnetic simulation of the distribution of magnetostatic fields $B_{demag}$ (arbitrary units) for (a) Fe(100)Au(x) and (b) Fe(200)Au(x) series of samples. (c) Parametric diagram of the dominating type of interactions represented by points obtained from simulations.

As Equation (1) shows, the type (I) interaction lowers $E_{ms}$ only in the case of the co-directional orientation of magnetization in adjacent Fe segments of the same BNW, favoring the uniform magnetization of the whole nanowire. In contrast, type (III) decreases $E_{ms}$ only in the case of the opposite direction of magnetization in Fe segments of neighboring BNWs, promoting the formation of the demagnetized array as a preferable state. Type (I) and type (III) interactions can be detected by the FORC-diagram method since they favor irreversible switching of the magnetization. The type (II) interaction does not induce any switching because it does not affect neighboring Fe segments, and consequently, it is not visible in the FORC diagrams.

## 3. Conclusion

This study carried out a detailed investigation of the interplay between the geometry and magnetic properties of BNW arrays. We found that the BNWs fabricated by electrodeposition from one solution consisted of the Fe-Au alloy with different concentrations of Fe and Au in the Fe and Au segments. Furthermore, the elapsed time corresponding to the reduction potential of each material affected the morphology of the nanowires, but not their microstructural modification. Based on this understanding, we deduce that morphological and geometrical features are decisive factors for manipulating the magnetic properties and micromagnetic structure of BNWs. Magnetic hysteresis measurements showed low coercive force and remnant magnetization. The MFM studies demonstrated that in the residual state, the BNWs are self-demagnetized, and in the presence of the external field, magnetic poles were induced at the ends of the Fe segments. Micromagnetic simulations allowed us to study the detailed micromagnetic structure and showed that nanowires in the remnant state break into vortex structures that alternate in space. Combining these results with the FORC-diagram method allowed us to estimate the effect of the segment size and composition on magnetic properties, particularly to describe the complex magnetostatic coupling between segments and nanowires in the array. We found that the observed magnetostatic interactions can be of three types: (I) intrawire interactions between adjacent Fe segments in the same BNW, (II) coupling of poles in the same Fe segment, and (III) interwire interaction between Fe segments in the neighboring BNWs in the array. The magnitude of all these interactions strongly depends on the geometrical parameters of the individual segments and BNWs array.

The results of this study could be helpful for the design and engineering of a magnetic memory cell and a logic device as well as for a better fundamental understanding of the magnetization confinement and magnetostatic interactions in a system of close-packed magnetic segments distributed in three-dimension space and of the influence of geometrical parameters of segments and a whole array on the type and magnitude of such interactions. The potential tunability of Fe/Au BNWs with the alternating magnetization makes them perfect applicants not only for next-generation nanoelectronics, but also for nanosensors, bio-barcoding, target drug delivery, biomedical assaying, and biomolecular separation.

## 4. Experimental Section

*Preparation of Fe-Au barcode nanowires:* Commercially available anodized aluminum oxide (AAO) membranes (Whatman Anodisc) with about 200 nm diameter of pores and 60 μm length were used as nano-templates in this study. A 300 nm Ag conducting layer was deposited on one side of the AAO membrane for nanowire electrodeposition. The electrolyte consisted of a mixture of deionized water and solutions of iron (II) sulfate heptahydrate ($FeSO_4 \cdot 7H_2O$, 0.10 M) as a source of Fe ions, and potassium dicyanoaurate ($KAu(CN)_2$, 0.015 M) as a source of Au ions. The pH level was controlled using boric acid ($H_3BO_3$, 0.80 M). Pulses with different current densities, 10 and 0.01 mA/cm$^2$, were fed into the electrochemical cell to grow Fe and Au segments, respectively. The growth rate for each segment was calculated as 14.6 and 1.65 nm/s for Fe and Au, respectively. After the nanowire synthesis, the conducting Ag layer was removed by an etchant. To investigate the geometric parameters, the elemental composition, and the micromagnetic structure, AAO was dissolved with sodium hydroxide (NaOH, 1 M) at 45°C for 10 min. After rinsing the prepared solution with deionized water several times to ensure a perfect removal of the residue, BNWs were collected using a magnet and moved to a Si or Cu substrate.

*Microstructural characterization:* BNW morphologies were studied with dual-beam scanning electron microscopy (SEM, ThermoFisher Scios 2 DualBeam) and ultrahigh-resolution field-emission SEM (UHR–SEM, Hitachi, SU–70). Furthermore, the microstructure was analyzed by transmission electron microscopy (TEM, JEOL, JEM–2100F) and X-ray diffraction (XRD, Rigaku, D/MAX–2500V/PC). The sample composition was probed by SEM energy dispersive X-ray spectroscopy (EDX, EDAX team).

*Magnetic properties:* Magnetic properties were studied using a vibrating sample magnetometer (LakeShore 7410 VSM). The FORC-diagram method was used to investigate the irreversible processes in the BNW arrays and the influence of interaction fields on magnetic behavior. Magnetic force microscopy (MFM, NT–MDT Ntegra Aura) was used to capture the micromagnetic structure of the BNWs.

*Micromagnetic Simulation:* Micromagnetic simulations were performed using the MuMax[3] software[23] to study different compositions of spin configurations in Fe-Au segments. The OOMMF,[30] ParaView, and MuView packages were used to visualize and analyze the simulated micromagnetic structure and magnetostatic field distribution.

*Statistical analysis:* For the statistical analysis of segment's composition (sample size $n = 50$, significance level $p = 0.05$), lengths (sample size $n = 50$, $p = 0.05$), and diameters of nanowires ($n = 75$, $p = 0.05$), independent two-sample *t*-test was used. Composition data were expressed as mean ± standard deviation.

**Supporting Information**

Supporting information is available from the Wiley Online Library or by direct contact with the author.

**Acknowledgments**

A. Y. S. and Y. S. J. contributed equally to this work. This research was supported by the National Research Foundation of Korea (2019R1A2C3006587). The Russian group thanks the Russian Ministry of Science and Higher Education for state support of scientific research conducted under the supervision of leading scientists in Russian institutions of higher education, scientific foundations and state research centers (075-15-2021-607).

**Conflict of Interest**

The authors declare that they have no known competing financial interests or personal relationships that could have appeared to influence the work reported in this paper.

**Data Availability**

Raw data were generated at Far Eastern Federal Univesity. Derived data supporting the findings of this study are available from the corresponding author Alexander S. Samardak on request.

**ToC Text**

We investigated the magnetic properties of metallic barcode nanowires (BNWs) composed of repeating Fe and Au segments fabricated by templated-assisted electrodeposition. We demonstrated that the dominating type of interaction depends on the geometric parameters of the Fe and Au segments and the interwire and intrawire distances.

**ToC Figure**

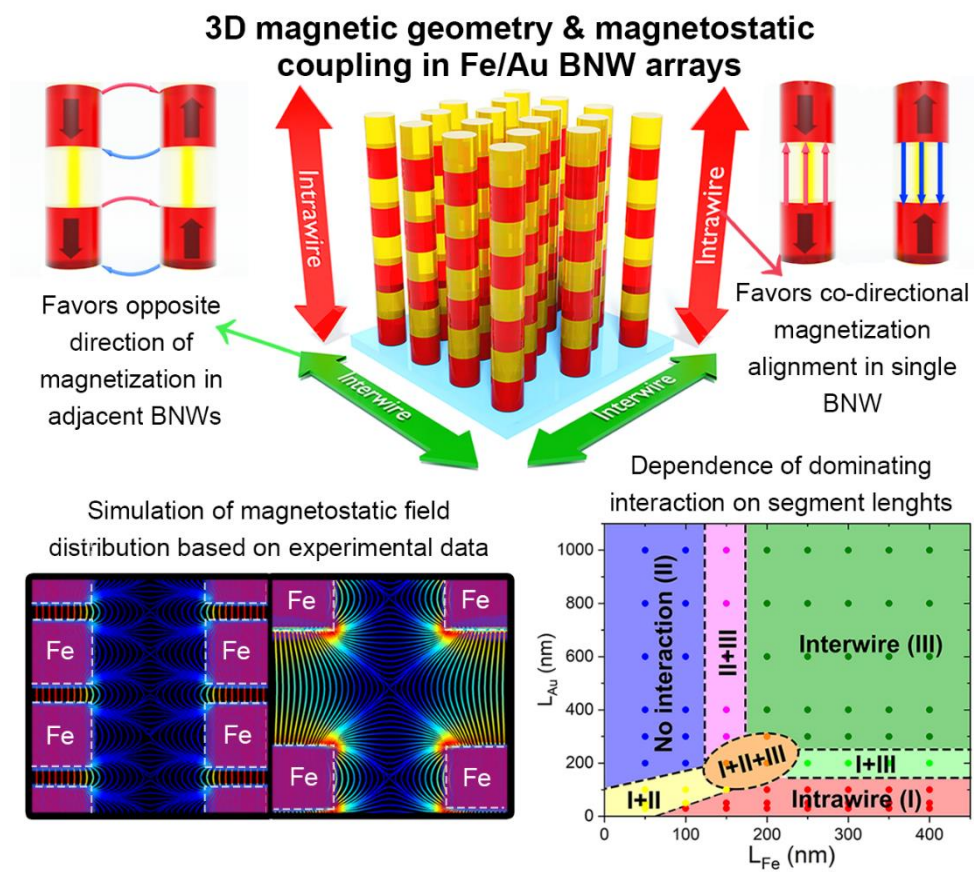

Supporting Information

# Interwire and Intrawire Magnetostatic Interactions in Fe-Au Barcode Nanowires with Alternating Ferromagnetically Strong and Weak Segments

*Aleksei Yu. Samardak, Yoo Sang Jeon, Vadim Yu. Samardak, Alexey G. Kozlov, Kirill A. Rogachev, Alexey V. Ognev, Eunjin Jeong, Gyu Won Kim, Min Jun Ko, Alexander S. Samardak,\* and Young Keun Kim\**

A. Y. Samardak, V. Y. Samardak, Dr. A. G. Kozlov, K. A. Rogachev, Dr. A. V. Ognev, Dr. A. S. Samardak
Department of General and Experimental Physics,
Far Eastern Federal University (FEFU), Vladivostok 690922, Russia
E-mail: samardak.as@dvfu.ru

Dr. Y. S. Jeon
Center for Hydrogen·Fuel Cell Research
Korea Institute of Science and Technology, Seoul 02792, Republic of Korea

E. Jeong, G. W. Kim, Dr. M. J. Ko, Prof. Y. K. Kim
Department of Materials Science and Engineering
Korea University, Seoul 02841, Republic of Korea
E-mail: ykim97@korea.ac.kr

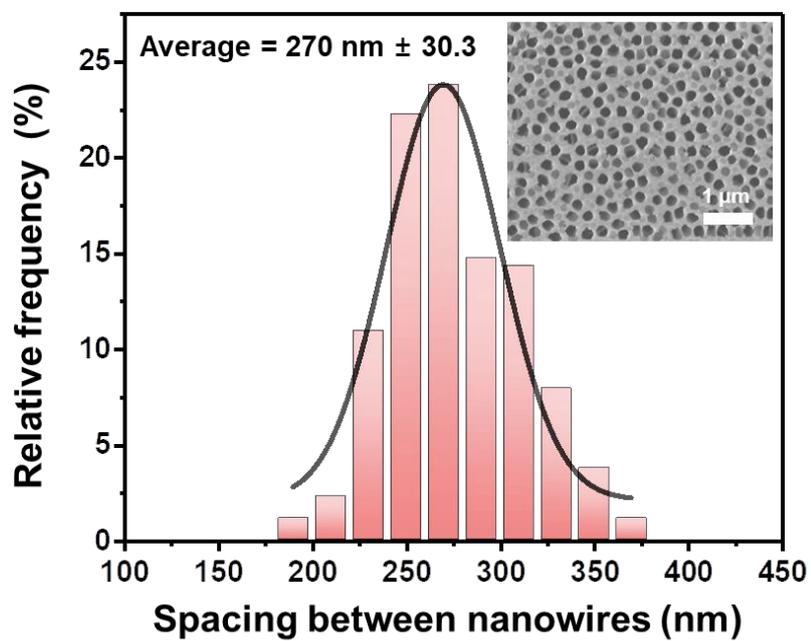

**Figure S1.** The spacing distribution between nanowires calculated from the top view image of the nanoporous AAO template (inset). The total $n = 265$ with $p = 0.05$.

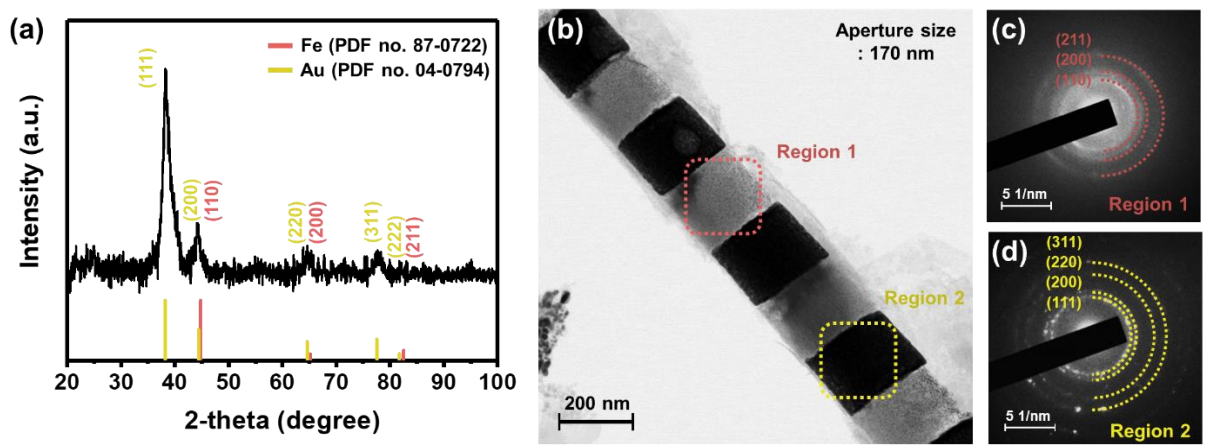

**Figure S2.** (a) XRD pattern indexed with PDF nos. 040794 (Fe) and 870722 (Au). (b) TEM image of a single Fe(200)Au(200) BNW and SAED patterns (c, d) obtained from Region 1 (Fe segment) and Region 2 (Au segment), respectively. Each sample showed the same trend regardless of the specification of the segment length.

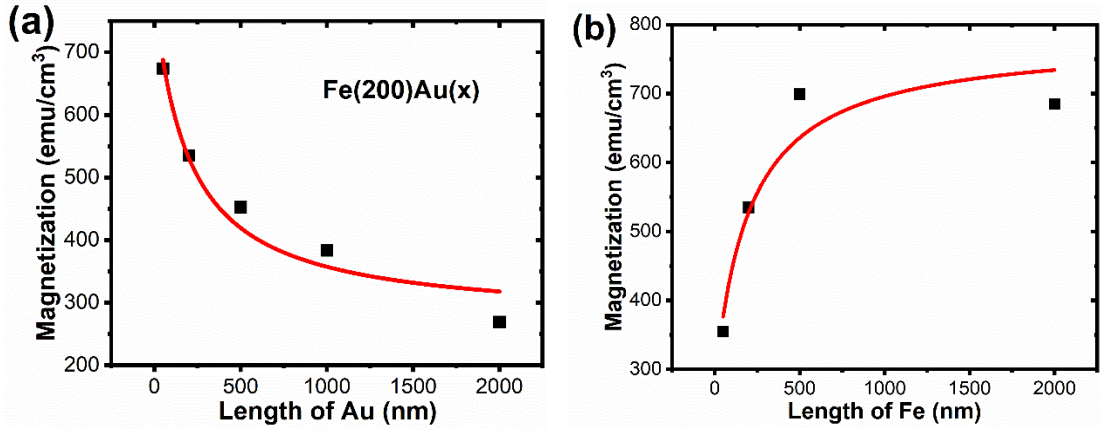

**Figure S3**. Magnetization acquired from different cases: (a) samples with constant $L_{Fe}$=200 nm, (b) samples with constant $L_{Au}$=200 nm.

The magnetization values were acquired from two different cases. The total values of the saturation magnetization values ($M_{s,tot}$) would be the sum of magnetization from the Fe segment ($M_{Fe}$) and the Au segment ($M_{Au}$) according to their ratio of the amount. The following equation served as the estimate where $h_{Fe}$ and $h_{Au}$ represented Fe and Au segment height, respectively.

$$M_{s,tot}=(M_{Fe}\,h_{Fe}+M_{Au}\,h_{Au})/(h_{Fe}+h_{Au}) \tag{S1}$$

The results obtained from (a) presented that $M_{Fe}$ and $M_{Au}$ were 800 emu/cm$^3$ and 254 emu/cm$^3$, respectively. Another case from (b) showed that $M_{Fe}$, and $M_{Au}$ were 882 emu/cm$^3$ and 216 emu/cm$^3$, respectively.

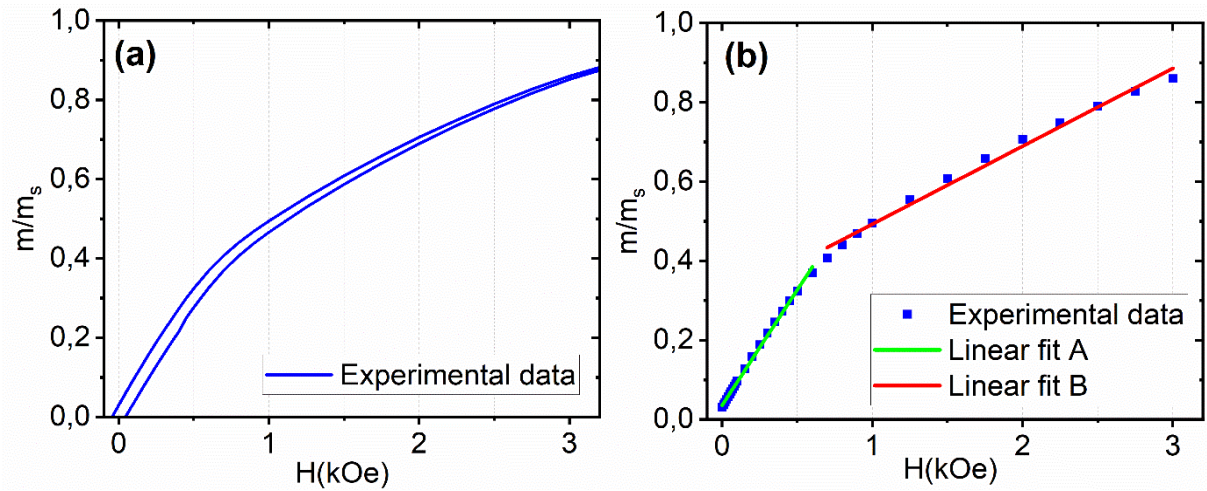

**Figure S4.** Close look at the inflection in the hysteresis loop for sample Fe(100)Au(200) and its decomposition into two linear fits. (a) Top right quarter of the experimental hysteresis loop for sample Fe(100)Au(200), obtained with an external field along the long axis of BNW. (b) Linear fits for two different regions of the same part of the corresponding hysteresis loop.

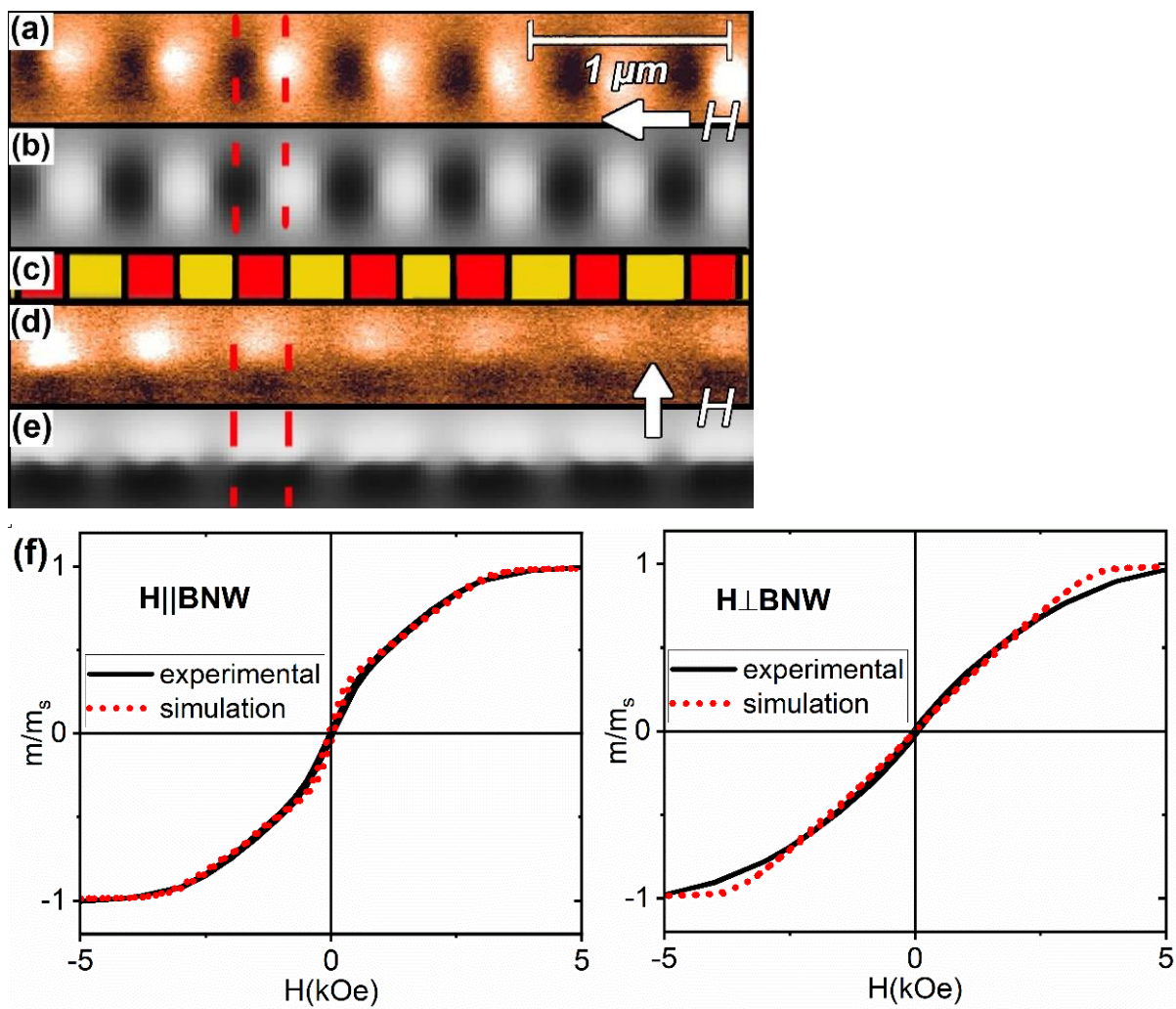

**Figure S5.** Comparison of simulated and experimental data (a) MFM contrast, $H$=700 Oe along with the long axis of BNW, (b) simulated stray fields, $H$ = 700 Oe along the long axis of BNW, (c) schematic of the corresponding segments, drawn based on the EDX results, red color exhibit Fe segment, yellow – Au, (d) MFM contrast, $H$ = 700 Oe perpendicular to the long axis of BNW, (e) simulated stray fields, $H$ = 700 Oe perpendicular to the long axis of BNW. (f) Comparison of simulated magnetic hysteresis loops with the experimental data for sample Fe(200)Au(200) for two directions of the external field. The dotted line shows the simulation results, while the solid line indicates the experimental data.

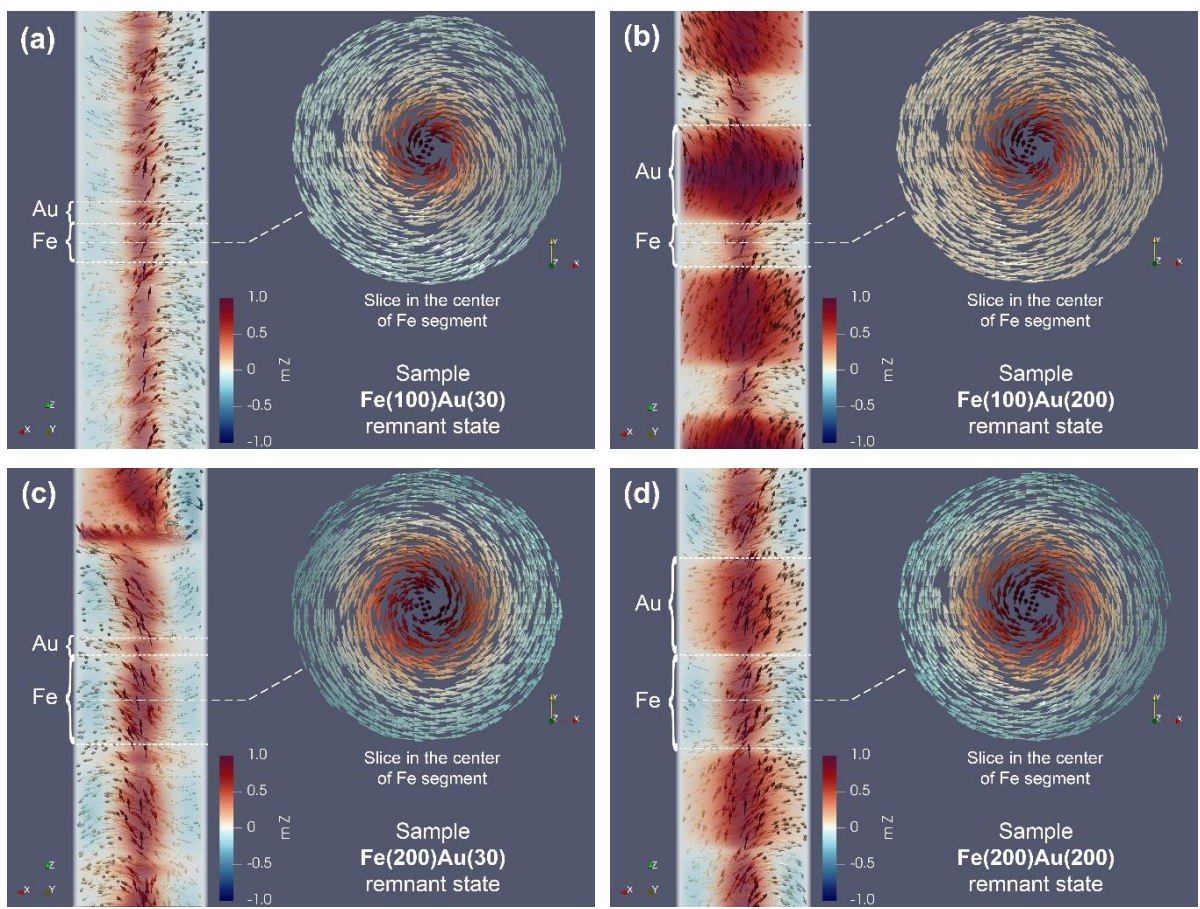

**Figure S6.** Simulated micromagnetic structure of (a) Fe(100)Au(30), (b) Fe(100)Au(200) (c) Fe(200)Au(30) and (d) Fe(200)Au(200) samples in the remnant state after saturation in the direction of the external magnetic field along the long axis of BNWs. The left side represents a slice in the center of a BNW, and the right side - a slice in the center of the Fe segment.

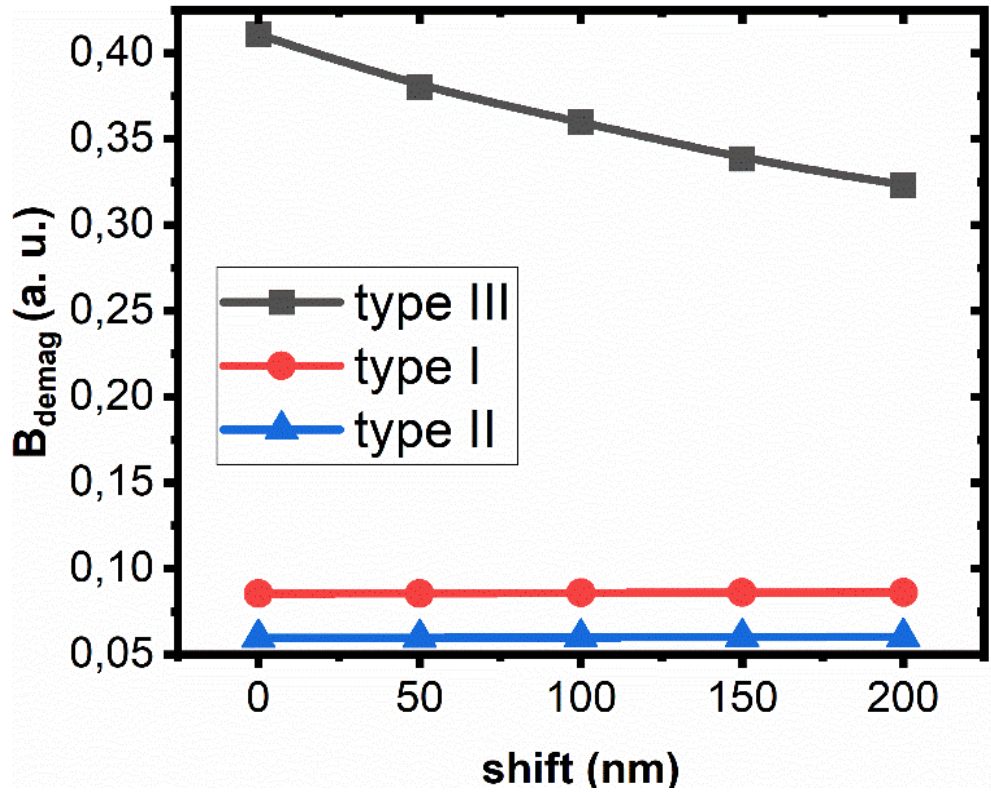

**Figure S7.** Dependence of $B_{\text{demag}}$ of three types of interactions in the system of two Fe(200)Au(200) BNWs shifted against each other along their length.

**Table S1.** The measured average length of Fe, Au segments, number of repeated cycles, and corresponding total lengths of each sample ($n = 50$ and $p = 0.05$ for individual samples).

| Sample name | Fe layer (nm) | Au layer (nm) | Repeat | Total length (µm) |
|---|---|---|---|---|
| Fe(100)Au(30) | 120.86 | 35.67 | 50 | **7.83** |
| Fe(100)Au(50) | 122.32 | 51.97 | 40 | **6.97** |
| Fe(100)Au(100) | 118.66 | 91.89 | 33 | **6.95** |
| Fe(100)Au(200) | 127.49 | 174.58 | 25 | **7.55** |
| Fe(200)Au(30) | 197.40 | 28.22 | 50 | **11.28** |
| Fe(200)Au(50) | 212.14 | 41.75 | 40 | **10.16** |
| Fe(200)Au(100) | 206.71 | 111.36 | 33 | **10.50** |
| Fe(200)Au(200) | 211.72 | 186.37 | 25 | **9.52** |

**Micromagnetic simulations in the applied field for samples:**

Movie 1. Fe(100)Au(30).

Movie 2. Fe(100)Au(200).

Movie 3. Fe(200)Au(30).

Movie 4. Fe(200)Au(200).

Movie 5. Animation showing the transformation of magnetostatic field distribution with the changing shift of nanowires for the Fe(200)Au(200) sample.